\documentclass[12pt,preprint]{aastex}

\slugcomment{2005, submitted to ApJ}

\shorttitle{TeV emission from the Galactic center} \shortauthors{Lu et al.}

\begin{document}


\title{Capture of a Red Giant by the Black Hole Sagittarius A* as a Possible
       Origin for the TeV Gamma Rays from the Galactic Center}
\author{Y. Lu\altaffilmark{1,2}, K. S. Cheng\altaffilmark{2}, Y. F. Huang\altaffilmark{3,2}}
\email{ly@bao.ac.cn}

\altaffiltext{1}{National Astronomical Observatories, Chinese
 Academy of Sciences, Beijing 100012, China}
 \altaffiltext{2}{Department of Physics, The University of Hong
 Kong, Pokfulam Road, Hong Kong, China}
 \altaffiltext{3}{Department of Astronomy, Nanjing University, Nanjing 210093, China}


\begin{abstract}
Non-thermal TeV $\gamma$-ray emission within a multiparsec has
been observed from the center region of our Galaxy. We argue that
these $\gamma$-rays are the result of transient activity of the
massive black hole Sgr A$^*$ that resides at the Galactic Center.
Several thousand years ago, the black hole may have experienced an
active phase by capturing a red giant star and forming an
accretion disk, temporarily behaving like an active galactic
nucleus. A powerful jet, which contains plenty of high speed
protons, was launched during the process. These runaway protons
interact with the dense ambient medium, producing TeV $\gamma$-ray
emission through the $\pi^0$-decay process. We show that the total
energy deposited in this way is large enough to account for
observations. The diffusion length of protons is also consistent
with the observed size of the TeV source.
\end{abstract}


\section{Introduction}
It is known that there are many remarkable high-energy sources
harbored in the Galactic center (GC) region \citep{Mel01}.
Recently, TeV $\gamma$-ray emission from the direction of the GC
has been reported by three independent groups, Whipple
\citep{Kos04}, CANGAROO (Collaboration of Australia and Nippon for
a Gamma Ray Observatory in the Outback; \citet{Tsu04}), and
HESS(High Energy Stereoscopic System; \citet{Aha04}). At least
four potential candidates are suggested for this TeV $\gamma$-ray
emission, which are the black hole Sgr $A^*$ \citep{Ahar05, Ato04,
Lev00}, the compact and powerful young supernova remnant (SNR) Sgr
A East \citep{Cro05}, the dark matter halo \citep{Hor05, pro05,
Gne04,Ell02}, and the whole diffusion {\rm 10\,pc} region
\citep{Aha05}. Interestingly, the angular scale of the TeV source
was determined by HESS to be less than a few arc-minutes,
indicating that this $\gamma$-ray source is located in the central
$\leq 10\,pc$ region \citep{Aha04}. It suggests that the black
hole Sgr $A^*$, with a mass of $2.6\times 10^6M_\odot$
\citep{Sch02}, should be involved \citep{Ahar05}.

Recently, a hadronic origin of TeV $\gamma$-rays that is linked to
the massive black hole has been addressed in detail by
\citet{Aha05}. They argued that the TeV $\gamma$-rays are produced
indirectly through the processes of $\pi^0$-decay when
relativistic protons are injected into the dense ambient gas
environment. The flux of this radiation component depends on the
density of the target, the diffusion speed of protons in the
interstellar medium, and the injection rate of protons. As a
result, a dense gas target and an extremely large proton flux is
required. However, since the black hole Sgr $A^*$ currently only
emits faint electromagnetic radiation, it is largely uncertain how
the protons could be accelerated to relativistic speeds.

It is well known that jets associated with accretion disks
surrounding black holes are efficient in accelerating particles.
For example, TeV $\gamma$-rays observed from several BL Lac
objects (a subclass of active galactic nuclei) are argued to
originate from relativistic jets \citep{Pia98}. The jet model is
also used to explain the production of TeV $\gamma$-rays from
microquasars \citep{bos05}. However, such a jet model cannot be
applied directly to the black hole Sgr $A^*$.  An extraordinarily
low bolometric luminosity of $\sim 10^{36}\,ergs\,s^{-1}$ has been
estimated for the black hole through multi-wavelength
observations, which indicates that Sgr $A^*$ is in its quiescent
dim state, and no powerful jet exists at present.

We propose that the black hole Sgr $A^*$ could be re-activated and
produce a powerful jet by capturing a red giant (RG) star,
temporarily behaving like an active galactic nucleus (AGN). This
may have happened thousands of years ago, naturally providing a
mechanism to generate the proton flux required by the hadronic
model of Aharonian \& Neronov. The structure of our paper is as
follows. The capture process is described in Sect.2. We estimate
the model parameters and the energy of the proton flux available
for the production of TeV $\gamma$-rays in Sect.3. A brief
discussion and conclusion are presented in Sect.4


\section{Red Giant Capture and the Lighting up of Sgr $A^*$ }
\subsection{the capture of a red giant}
Many if not all galaxies are expected to posses a massive black
hole \citep{Hae93}. It has been thought that the capture and tidal
disruption of stars by such a massive black hole could result in
flarelike activities in the central regions of AGNs, and even in
normal galaxies and globular clusters. The flare results from the
rapid release of gravitational energy as the matter from the
disrupted star plummets toward the black hole \citep{Ree90,
Beg90}. This process may even be the mechanism of $\gamma$-ray
bursts \citep{Che01}. It is interesting to note that the presence
of an extended high-temperature plasma at the GC region has been
confirmed by Advanced Satellite for Cosmology and Astrophysics
(ASCA) observations \citep{Koy96}, suggesting that the GC
exhibited intermittent activities at least 300 yr ago. We propose
that the black hole Sgr $A^*$ could have been naturally lit up and
produced the currently observed TeV $\gamma$-rays when it captured
a RG star.

The detailed capture and disruption process of a main-sequence
(MS) star has been studied by several authors \citep{Ree88,
Can90}. Syer \& Ulmer (1999) also discussed the capture rate of
red giant stars by a massive galactic black hole in virialized
star clusters. Compared with the capture of a MS star, the
disruption of a RG star is different in that (1) RG disruption
events generally last much longer, but are relatively fainter; (2)
the tidal disruption radius of a RG star is typically equal to its
pericenter radius; and (3) the size of the accretion disk formed
in the capture process of a RG is larger. We believe that the
capture of a RG star (rather than a MS star) is more appropriate
to explain the observed $\gamma$-ray emission from the GC (this
problem is further addressed in section 4).

We follow the treatments of Syer \& Ulmer (1999) to study the
capture of a RG star and scale our parameters to the physical
conditions in the Galactic center. When a star of mass $M_*$ and
radius $R_*$ passes by a black hole with a mass of $M_{bh}$, the
star could be captured at an average tidal radius of
\begin{eqnarray}
R_T\approx R_*(\frac{M_{bh}}{M_*})^{1/3}\simeq 8.35\times 10^{13}
M_6^{1/3}m_*^{-1/3}r_*\,\,cm\,\,\,,
\end{eqnarray}
where $m_*=M_*/M_\odot$, $M_6=M_{bh}/10^6M_\odot$, and
$r_*=R_*/12R_\odot$. For MS stars, $R_*=1R_{\odot}$, but for RG
stars, the radius ranges from $3$ to $200 R_\odot$ \citep{Sye99}.
The maximum radius of a RG star ($200\,R_\odot$) is attained only
during a short period ($10\%$) of its evolutionary lifetime, so
the typical radius of RG stars is actually $\sim 12R_\odot$, and
we take $r_*=1$ in our model. The capture rate of MS stars by a
massive black hole has been discussed by Phinney (1989), Rees
(1990), and Cannizzo et al.(1990). However, the capture rate of RG
stars is more complicated, involving the mass function of stars,
stellar evolution, the black hole mass, and the maximum radius of
a RG \citep{Sye99}. By assuming a Salpeter mass function for the
stars, Syer \& Ulmer (1999) estimated the capture rate in our
Galaxy as $\dot{N}_{MS}\sim 4.78\times 10^{-5}\,\,yr^{-1}$ and
$\dot{N}_{RG}\sim 8.51\times 10^{-6}\,\,yr^{-1}$ (Table 1 in Syer
\& Ulmer 1999) for MSs and RGs, respectively.

Once a RG star is tidally disrupted, a fraction of its debris
plummets to the black hole. The time for the debris to return to
the pericenter ($t_{min}$) and the time for it to enter a circular
orbit around the black hole ($t_{cir}$ ) are \citep{Ulm99}
\begin{eqnarray}
t_{min}& =& \frac{2\pi R^3_T}{(GM_{bh})^{1/2}(2R_*)^{3/2}} \approx
4.57 m_*^{-1}r_*^{3/2}M_6^{1/2}\,\,yr\,\,,
\end{eqnarray}
 \begin{eqnarray}
t_{cir} &=& n_{orb}t_{min}\approx
9.15(\frac{n_{orb}}{2})(\frac{t_{min}}{4.57\,yr})\,\,yr\,\,\,,
\end{eqnarray}
respectively, where $n_{orb}$ is the number of orbits necessary
for circularization, probably ranging between 2 and 10. We assume
that the pericenter radius of the captured star equals its capture
radius \citep{Sye99} in the calculations.

After the debris circularization, either a thick torus or a thin
disk would be formed around the black hole, depending on two
timescales. One is the time to radiate a significant portion of
the energy of the bound debris at the Eddington limit ($t_{rad}$),
the other is the time to accrete a large fraction of the disk
($t_{acc}$). If $t_{acc} \gtrsim t_{rad}$, a thin disk will form;
otherwise, a thick torus will appear \citep{Ulm99}. The former
case is necessary to our model, since the jet associated with the
accretion disk is efficient to accelerate particles. Below we show
that a thin disk will truly be formed in the process that we
consider.

The radiation timescale is \citep{Sye99, Ulm99}
\begin{eqnarray}
t_{rad}& = &\frac{\xi m_*M_\odot c^2\eta}{L_Edd}\approx
21\xi_{0.5} m_*M_6^{-1}\eta_{0.1}\,\,yr\,\,\,,
\end{eqnarray}
where $\xi=0.5\xi_{0.5}$ is the fraction of the stellar mass that
goes into the bounded debris \citep{Ree88}, $L_{Edd}$ is the
Eddington luminosity, and $\eta=0.1\eta_{0.1}$ is the efficiency
of transferring the rest mass to radiation energy in the accretion
process, which is defined as $\eta=L_{disk}/\dot{M}c^2$, where
$L_{disk}$ is the total radiated disk luminosity, and $\dot{M}$ is
the mass accretion rate of the disk. For a rapidly rotating black
hole, $\eta \sim 0.42$.

For a standard accretion disk  \citep{Sha73}, the accretion time
can be estimated as,
\begin{eqnarray}
t_{acc}=(\frac{h}{r})^{-2}\alpha^{-1}\Omega_K^{-1}\approx
2.08\times 10^4
(\frac{h}{r})_{-2}^{-2}\alpha_{-3}^{-1}m_*^{-1/2}r_*^{3/2}\,\,\,yr\,\,\,,
\end{eqnarray}
where $r$ and $h$ are the radius and height of the disk,
respectively, $\alpha$ is the viscous constant, and
$\Omega_K=(GM_{bh}/R_T^3)^{1/2}$ is the Keplerian velocity. The
typical values for h/r are in the range $10^{-2}$ to $10^{-1}$
\citep{Sha73}. We define $(h/r)_{-2}=(h/r)/10^{-2}$ and
$\alpha_{-3}=\alpha/10^{-3}$. From Eqs.(4) and (5), we find
$t_{acc}\gg t_{rad}$, thus a thin disk forms in our model.

The lifetime of our Galaxy is $t_{G}\sim 10^{10}\,yr$. We find
that when the black hole captures a MS or a RG star, the relation
of $\max(t_{rad}, t_{cir}, t_{acc}) < \min(1/\dot{N},\, t_{G})$ is
satisfied, where $\dot{N}={\dot{N}_{MS}\; or\; \dot{N}_{RG}}$.
This implies that the activity of the black hole triggered by the
capture events is intermittent. For the RG star capture, one long
flare can be observed that lasts for $t_{acc}\sim 2.08\times
10^4\,yr$.


\subsection{Accretion}

The time that the black hole becomes bright and active after the
RG star capture depends on the evolution of its surrounding
accretion disk. The overall evolution of the accretion rate is
illustrated in Fig.1. Initially, the accretion rate of the disk
exhibits an abrupt increase, since a local mass clumping in the
inner-most orbit of the disk is similar to a delta-function (Rees
1998; see Fig.1). This stage will last until the mass accumulation
rate of the disk reaches its peak $\dot{M}_{peak}$. We refer to
this stage as phase 1. According to the simulations of Evans \&
Kochanek (1989), the value of $\dot{M}_{peak}$ is
\begin{eqnarray}
\dot{M}_{peak}\sim 1.36\eta_{0.1}
m^2_*r^{-3/2}_*M_6^{-3/2}\dot{M}_{Edd}\,\,,
\end{eqnarray}
where $\dot{M}_{Edd}=2.46\times 10^{-2}M_6\eta_{0.1}^{-1}\,
M_\odot\,yr^{-1}$ is the Eddington accretion rate.

When $t>t_{peak}$, the accretion rate evolves as \citep {Ree88,
Phi89}
\begin{eqnarray}
&& \dot{M}\sim
\frac{1}{3}\frac{M_*}{t_{min}}(\frac{t}{t_{min}})^{-5/3}\,\,\,,\nonumber\\
&&\sim 2.97\eta_{0.1}
m^2_*r^{-3/2}_*M_6^{-3/2}\dot{M}_{Edd}(\frac{t}{t_{min}})^{-5/3}\,\,.
\end{eqnarray}
From this point on, the black hole is activated and enters a
luminous phase, acting like an AGN. This very active phase will
continue until a critical time ($t_{crit}$) when the actual
accretion rates falls below a critical value $\dot{M}_{crit}$
(Beckert \& Duschl 2004; 2002). We refer to this stage as phase 2.
During this phase, the radiation efficiency of the accretion disk
is very high. After phase 2, the accretion rate itself still
continues to fall slowly as $\propto t^{-5/3}$, but the accretion
flow becomes advection dominated. When advection takes over, the
radiation efficiency drops drastically by several orders of
magnitude (at least $\sim 10^{3}$; Beckert \& Duschl (2002)),
resulting in a switch off of the AGN-like phase. We describe this
stage as phase 3, during which the black hole becomes dim and no
longer looks like a quasar. Beckert \& Duschl (2002) pointed out
that $\dot{M}_{crit}$ should be less than $3\times
10^{-3}\dot{M}_{Edd}$ because no consistent advection-dominated
accretion flow (ADAF) models with $\dot{M}> 3\times
10^{-3}\dot{M}_{Edd}$ are feasible. However, the exact value for
$\dot{M}_{crit}$ is still uncertain, since it is related to the
radiation mechanism that is taking effect during the phase
transition \citep{Bla99, Qua99}. In our study, we use the observed
data to determine the value of $\dot{M}_{crit}$. Assuming that the
transition from phase 2 to phase 3 is triggered by thermal
instability of the disk \citep{Sha73, Lig74, Lu00}, the
corresponding transition timescale is
\begin{eqnarray}
\Delta t_{tr}\sim (\alpha\Omega_K)^{-1}\approx 2.08
\alpha_{-3}^{-1}m_*^{-1/2}r_*^{3/2} \,\,yr\,.
\end{eqnarray}
We see that the transition process is finished almost instantly.
Note that a faint X-ray source \citep{Bag03} has been
identified at the position of the GC, whose infrared radiation is
also weak \citep{Gen03}. It indicates that the present accretion
disk surrounding the black hole Sgr $A^*$ is probably ADAF
\citep{Nar95, Esi98, Fal00}. These observations also support our
idea that the black hole is currently in phase 3.

The time corresponding to $\dot{M}_{peak}$ can be estimated from
Eq.(7) as
\begin{eqnarray}
t_{peak} & \simeq & 1.59t_{min} \approx 7.25
m_*^{-1}r_*^{3/2}M_6^{1/2}\,\,yr\,\,.
\end{eqnarray}
Although the jet formation itself is very difficult to model,
observations are providing increasing evidence for the existence
of powerful collimated outflows in black hole systems
\citep{Bla79, Fen05, Gal05, Hei05}. We assume that when the black
hole Sgr $A^*$ evolves from phase 1 to phase 2, a powerful jet
would form, originating at the inner edge of the disk
\citep{Mar01, Che01, Lu00, Lu03}. For convenience, we define a
dimensionless parameter
\begin{eqnarray}
q_j=\frac{\dot{Q}_j}{\dot{M}c^2}\,,\,\,
\end{eqnarray}
where $\dot{Q}_j$ is the total jet power, including the rest
energy of the expelled matter. The value of $q_{j}$  can range
from $6\times 10^{-3}$ to $0.4$ \citep{Fen05, Gal05, Hei05}.

It has been suggested that most of the black holes in the universe
are rapidly spinning \citep{Elv02, Vol05}. Recently, by analyzing
the light curves of the X-ray and infrared flares from the GC
region, the angular momentum of the black hole Sgr $A^*$ has been
accurately determined as $a=0.9939^{+0.0026}_{-0.0074}$
(Aschenbach et al. 2004). The possible mechanism for spinning up
this black hole has been discussed by Y. Liu et al. (2006).
Therefore, for Sgr $A^*$, the radiation efficiency may be as high
as $\eta\approx 0.42$. Combining Eqs.(7) and (10), we have
\begin{eqnarray} &&
\dot{Q}_j = 1.75\times 10^{44}
\eta_{0.42}q_{j,-2}m^2_*r^{-3/2}_*M_6^{-1/2}(\frac{t}{t_{min}})^{-5/3}\,\,ergs\,s^{-1}\,,
\end{eqnarray}
where $\eta_{0.42}=\eta/0.42$ and $q_{j,-2}=q_j/10^{-2}$. The
total energy of the jet is then
\begin{eqnarray}
E_{j,tot}=\int^{t_{crit}}_{t_{peak}}\dot{Q}_{j}dt\,\,\,.
\end{eqnarray}
The value of $t_{crit}$ is determined in the next section.

\section{Jet Energy and TeV Gamma Ray Emission}

In spite of extensive observations, direct evidence for
high-energy activities of the GC region, such as those observed in
AGNs, is still lacking. Fortunately, strong fluorescent X-ray
emission has been found from the cold iron atoms in the molecular
cloud at the Sgr B2 region (Sunyaev et al. 1993; Koyama et al.
1996). The most favorable explanation for this emission is that
the molecular cloud was once illuminated by intense X-rays from
Sgr A$^*$ (Sunyaev \& Churazov 1998; Cramphorn \& Sunyaev 2002).
According to this theory, the X-ray luminosity of Sgr A$^*$ should
have been $L_X \sim 2 \times 10^{39} ergs \; s^{-1}$ about 300 yr
ago, which was most likely due to the capture of a star by the
black hole. However, observations also show that the current X-ray
luminosity of Sgr A$^*$ is only $2 \times 10^{36} ergs \; s^{-1}$
(Koyama et al. 1996). Therefore, if this explanation is correct,
then the luminosity of Sgr A$^*$ has decreased rapidly by a factor
of $\sim 10^3$ in less than 300 yr. We note that in our framework,
such a rapid dimming is possible only at the time of $t_{crit}$,
when the accretion disk formed by capturing a star transitions
from phase 2 to phase 3. As stated in Section 2.2, the transition
could be finished in a few years (see Eq.(8)), and the luminosity
could decrease by a factor of $\sim 10^3$ after the transition.
Thus, we suggest that the required luminosity of $L_X \sim 2
\times 10^{39} ergs \; s^{-1}$ was just the power of the black
hole system when it was approaching the end-point of phase 2. Then
the lifetime of the jet can be determined by setting $\eta \dot{M}
(t_{crit}) c^2 = L_X$. The result is
\begin{eqnarray}
t_{crit}\sim 1.68\times
10^4\eta_{0.42}^{3/5}m_*^{-1/5}r_*^{3/5}M_6^{1/5}\,yr\,\,.
\end{eqnarray}
Combining Eqs.(9)and (11)-(13) and taking $\eta=0.42$,
$q_j=10^{-2}$,  we obtain
\begin{eqnarray}
E_{j,tot}
 \simeq 2.76\times 10^{51}\,ergs\,\,.
\end{eqnarray}
If a larger value of $q_j$ is assumed, then the $E_{j,tot}$ would
be even higher. Note that $E_{j,tot}$ is the total kinetic energy
of the jet. Possibly only $\sim 10\%$ of it could be converted
into high-energy particles.

The energy of the injected protons required by the hadronic origin
of TeV $\gamma$-rays has been modeled by Aharonian \& Nerono
(2005b). It depends on the diffusion coefficient of protons in the
ambient gas. Generally, the dependence of the diffusion
coefficient on proton energy can be expressed as $D(E)=
10^{28}(E/1GeV)^\delta\kappa\,cm^2\,s^{-1}$, where $\delta\sim
0.3$-$0.6$ \citep{Ber90} and $\kappa\sim 10^{-4}$ to $10^{-2}$
\citep{Aha96} are dimensionless parameters. There are mainly three
typical propagation scenarios, characterized by different $\delta$
and $\kappa$ values \citep{Aha05}. The case of the effective
confinement of protons (ECP) corresponds to $\delta=0.5$ and
$\kappa=10^{-4}$, the Kolmogorov type turbulence (KTT) corresponds
to $ \delta=0.3$ and $\kappa=0.15 $, and the Bohm diffusion (BD)
corresponds to $\delta\sim 1.0$ and $\kappa\sim 10^{-2}$. For
photons with a typical energy of 10 TeV, the diffusion coefficient
is $1.0\times 10^{26}\,\,cm^2\,s^{-1}$ for ECP, $2.4\times
10^{28}\,\,cm^2\,s^{-1}$ for KTT, and $1.0\times
10^{30}\,\,\,cm^2\,s^{-1}$ for BD.

Assuming that the injection of protons lasts for $\sim 10^5\,yr$,
the required injection energy by the hadronic origin of the TeV
$\gamma$-rays in these three cases can be estimated as
\begin{eqnarray}
E_{jtot}=\dot{W}_p10^{5}\,yr &\approx& \left\{\begin{array}
{r@{\;,\quad}l}
 2.21\times
10^{49}\,\,ergs\,& {\rm for\; ECP\;}
 \,\,\,,\\
 2.38\times 10^{50}\,ergs\,\,& {\rm for\;KTT\;}
 \,\,\,,\\
 3.15\times
10^{51}\,ergs\,\,& {\rm for\;BD\;}
 \,\,\,,
 \end{array}
\right.\,
\end{eqnarray}
where $\dot{W}_P$ is the required injection rate of protons
corresponding to the three diffusion mechanisms, which is
\citep{Aha05},
\begin{eqnarray}
\dot{W}_p &\approx& \left\{\begin{array} {r@{\;,\quad}l}
 7.0\times
10^{36}\,\,ergs\,s^{-1}\,& {\rm for\;ECP\; \;}
 \,\,\,,\\
 7.5\times 10^{37}\,\,ergs\,s^{-1}\,& {\rm for\;KTT\;}
 \,\,\,,\\
 1.0\times
10^{39}\,\,ergs\,s^{-1}\,& {\rm for\;BD\;}
 \,\,\,.
 \end{array}
\right.\,\,
\end{eqnarray}
Comparing Eq.(15) with Eq.(14), and taking into account the
efficiency ($\sim$ 10\%) of converting $E_{j,tot}$ into cosmic
rays, we find that the required energy could be satisfactorily
provided by our RG star capture model in at least two diffusion
cases, i.e., the ECP and KTT diffusion. In fact,
 if we assume $q_j > 0.1$, then the jet energy would also be
enough for the BD mechanism.

Note that the linear size of the TeV source is likely to be only
$\sim 10\,pc$ (Aharonian et al. 2004). We now further check which
propagation scenario can meet this requirement. Theoretically, the
protons would diffusion to a radius of
$R=[4D(10\,TeV)t_{dif}]^{1/2}$ \citep{Aha96}, where $t_{dif}$ is
the diffusion time of the injected protons when they propagate in
the target. In our case, $t_{dif}=t_{crit}+300\,yr$. For the three
diffusion mechanisms, the radius of the source is,
\begin{eqnarray}
R=\sqrt{4D(10\,TeV)t_{dif}} &\approx& \left\{\begin{array}
{r@{\;,\quad}l}
 4.89\,pc& {\rm for\; ECP\;},
 \,\,\,\\
 75.9\,pc & {\rm for\;KTT\;},
 \,\,\,\\
 489 \,pc & {\rm for\;BD\;}.
 \,\,\,
 \end{array}
\right.
\end{eqnarray}
Eq.(17) shows that for both BD and KTT propagation scenarios, the
corresponding radius is obviously too large as compared with
observation. On the contrary, in the case of the ECP propagation
scenario, we obtain $R\sim 4.89\,pc$, then the linear size is
$2R=9.78\,pc$. It is in good agreement with the observed size of the TeV
source. We thus believe that the ECP diffusion is taking effect in
our framework.

In short, our analysis shows that the capture of a RG star by the
black hole Sgr $A^*$ could be the energy source of the observed
TeV $\gamma$-ray emission from our GC. The energy deposited in
this way is large enough to produce a relativistic outflow of
protons. The outflow could diffuse into a volume with the diameter
of $\sim 10\,pc$. The detailed process that transfers the energy
of the injected protons into that of observed $\gamma$-rays has
been discussed by other authors (Aharonian et al. 2004; Aharonian
\& Neronov 2005b), and is not addressed repeatedly here.


\section{Discussion and Conclusions}

A strong TeV $\gamma$-ray source has been detected at the Galactic
center,  whose size is probably less than 10 pc. It should be
closely related to the black hole Sgr A$^*$ of our Galaxy.
Aharonian \& Neronov (2005b) suggested a hadronic origin for these
$\gamma$-rays; i.e., they are produced through $\pi^0$-decay
process when a strong flow of relativistic protons interacts with
the dense ambient gas. However, the nature of this proton flow is
largely unknown. In this paper, we show that the required proton
outflow could have been reasonably produced when the black hole
Sgr A$^*$ captured a RG star and formed an accretion disk around
it in the past. The whole process can be divided into three
phases. In phase 2, which lasts for $\sim 10^4$ yr, the black hole
becomes active and luminous thanks to a relatively high accretion
rate, temporarily behaving like an AGN. A relativistic outflow of
protons can be ejected in this phase, whose total energy is as
high as $E_{j,tot} \sim 2.76 \times 10^{51}$\,ergs, large enough
to meet the requirement of the Aharonian \& Neronov model.

According to our calculations, the propagation of protons in the
target gas is through the ECP scenario. We show that the injected
protons have diffusiond into a volume of $\sim 9.78\,pc$,
consistent with the observationally inferred size of the TeV
source. Furthermore, when the 10 TeV protons diffuse in the target
gas, their escape time is comparable with the characteristic time
of proton-proton interactions, implying that the spectrum of the
observed TeV $\gamma$-rays should be similar to that of the
injected protons \citep{Aha05}. Since the $\gamma$-ray spectrum
observed by HESS is, $J(E)=(2.5\pm0.21)\times 10^{-12}E^{-
2.21}\,photon\,(cm^2\,s\,TeV)^{-1}$ \citep{Aha04}, we can
conjecture that the initial spectrum of the injection protons
should be $Q(E) \propto E^{-2.2}\exp({-E/E_0})$. Aharonian \&
Neronov (2005b) pointed out that the cutoff energy is
$E_0=10^{15}\,eV$.

In our study, we mainly consider the case in which the captured
star is a RG star. If the captured star is a MS one, things will
be much different. In Table 1, we have listed the key quantities
calculated for a MS star capture, comparing them directly with
those of a RG star capture. Interestingly enough, we find that the
energy injection by a MS star capture also meets the requirement
of energetics. However, we note that $t_{crit}>t_{acc}$ for a MS
star; this means that the jet formed by the capture of a MS star
terminates much earlier. In this situation, the diffusion
timescale of the injected protons is $t_{dif}\sim t_{acc}+
300\,yr$, which is much smaller than the diffusion time of a RG
capture event. Consequently, within the jet lifetime, the protons
originating from a MS star capture will diffusion to
\begin{eqnarray}
R=\sqrt{4D(10\,TeV)t_{dif}} &\approx& \left\{\begin{array}
{r@{\;,\quad}l}
 1.05\,pc& {\rm for\;the\; ECP\;},
 \,\,\,\\
 16.4\,pc & {\rm for\;the\;KTT\;},
 \,\,\,\\
 105 \,pc & {\rm for\;the\;BD\;},
 \,\,\,\nonumber
 \end{array}
\right.
\end{eqnarray}
which is inconsistent with the size ($\sim 10$ pc) of the TeV
source inferred from current observations. This is the main reason
that we prefer a RG star capture rather than a MS star capture in
our framework.

However, to determine the size of the TeV source observationally
is not an easy task because of its extended nature. We note that
the HESS collaboration has not yet published the final analysis of
its observations of the Galactic center on larger scales, so there
is still a lack of information on TeV emission beyond the 10 pc
scale at the area. It is thus possible that the actual size of the
TeV source may be larger. In addition, the reduction of TeV
emission may also be due to the decrease of the density of the gas
at larger distances, but not the lack of an ultra-relativistic
proton flow. Taking into account these factors, the MS star
capture scenario still cannot be completely excluded. In fact, a
comprehensive analysis of the TeV radiation and a thorough
investigation of the environment within $\sim 100$ pc around Sgr
A$^*$ are necessary for us to understand the history of the
activities in the GC region.

We thank the anonymous referee for valuable comments and
suggestions that lead to an overall improvement of this study. We
are grateful to S. N. Zhang for very valuable discussions and
thoughtful comments. Thanks also goes to W. Wang for helpful
discussions. This research was supported by a RGC grant of Hong
Kong government, by the National Natural Science Foundation of
China (Grants 10273011, 10573021, 10433010, 10233010, and
10221001), by the Special Funds for Major State Basic Research
Projects, and by the Foundation for the Author of National
Excellent Doctoral Dissertation of P. R. China (Project No:
200125.

\clearpage
\begin{table}
\begin{center}
\caption{Key quantities for a MS star capture and a RG star
capture .\label{tbl-1}}
\begin{tabular}{cllllll}
\tableline\tableline Star&$r_*$ & $t_{min}(\,yr)$ &
$t_{peak}(\,yr$) & $t_{acc}(\,yr)$ & $t_{crit}(\,yr)$ &
$E_{jtot}(\,ergs)$
 \\
\tableline
MS & $1/12$&0.11 &0.174 &$5.0\times 10^2$ &$3.78\times10^3$ &$ 2.76\times 10^{51}$ \\
RG & 1&4.57 &7.25 & $2.08\times10^4$&$1.68\times10^4$ &$ 2.76\times 10^{51}$\\
\tableline
\end{tabular}


\end{center}
\end{table}
\clearpage
\begin{figure}
 \epsscale{0.80}
 \plotone{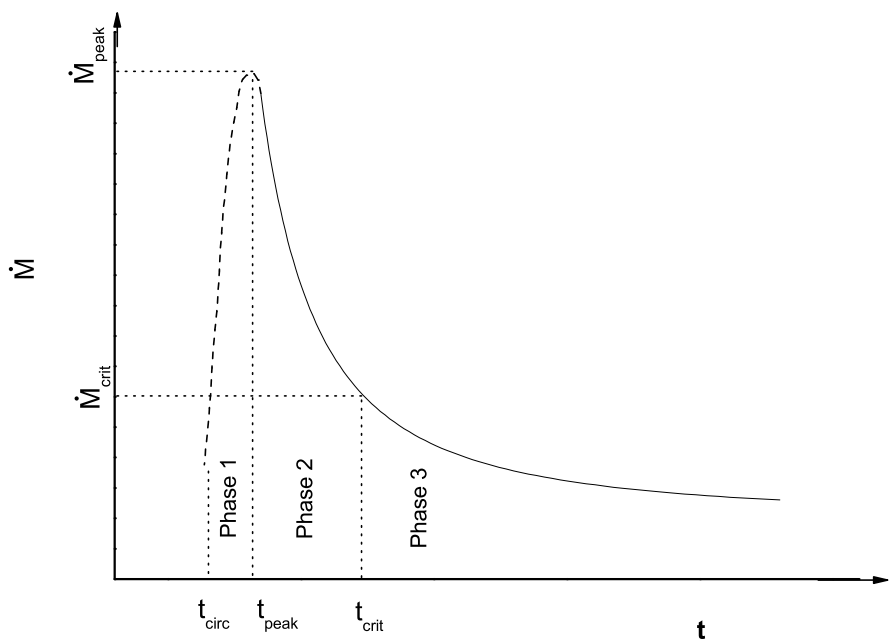}
 \caption{The behavior of the accretion rate versus time ($\dot{M}-t$) when a black hole
 captures a star. The dashed line represents the local mass clumping
 in the inner region of the disk. The solid line is the evolution of the accretion
 rate as a function of $t^{-5/3}$.
  \label{fig1}}
\end{figure}

\end{document}